\documentclass{aip-cp}

\usepackage[numbers]{natbib}
\usepackage{rotating}
\usepackage{graphicx}
\usepackage{amsmath1}
\newcommand{\be}{\begin{equation}}
\newcommand{\ee}{\end{equation}}
\newcommand{\bea}{\begin{eqnarray}}
\newcommand{\eea}{\end{eqnarray}}
\newcommand{\ba}{\begin{array}}
	\newcommand{\ea}{\end{array}}

\begin{document}

% Title portion
\title{
	Predictions for Diffractive $\phi$ Meson Production Using an AdS/QCD Light-front Wavefunction}

\author[aff1]{Mohammad Ahmady\corref{cor1}}
%\eaddress[url]{http://www.aip.org}
\author[aff1,aff2]{Ruben Sandapen}
\eaddress{ruben.sandapen@acadiau.ca}
\author[aff3]{Neetika Sharma}
\eaddress{neetika@iisermohali.ac.in}

\affil[aff1]{Department of Physics, Mount Allison University, Sackville, New Brunswick E4L 1E6, Canada}
\affil[aff2]{Department of Physics, Acadia University, Wolfville, Nova Scotia B4P 2R6, Canada}
\affil[aff3]{Department of Physical Sciences, Indian Institute of Science Education and Research Mohali,
	S.A.S. Nagar, Mohali-140306, Punjab, India}
\corresp[cor1]{Corresponding author: mahmady@mta.ca}

\maketitle

\begin{abstract}
We compute the rate for diffractive $\phi$ electro-production using the Color Glass Condensate dipole model.  The model parameters are obtained from fits to the most recent combined HERA data on inclusive deep inelastic scattering.  As for the $\phi$ meson, we use an AdS/QCD holographic light front wavefunction. Our predictions are compared to the available data collected at the HERA collider.
\end{abstract}

% Head 1
\section{INTRODUCTION}
The light-front wavefunction (LFWF) of the vector meson is an input in the QCD colour dipole model for calculating diffractive vector meson production.  In Ref. \cite{Forshaw:2012im}, successful predictions were obtained for diffractive $\rho$ production using a holographic wavefunction for the $\rho$ meson. The holographic meson wavefunction is predicted in holographic light-front  QCD proposed by Brodsky and de T\'eramond recently reviewed in \cite{Brodsky:2014yha}. In this work, we first use the new deep inelastic scattering (DIS) data from HERA collider, which were released in 2015 \cite{Abramowicz:2015mha}, to update the parameters of the color glass condensate (CGC) dipole model and then make predictions for diffractive $\phi$ meson production when a holographic LFWF is assumed for this vector meson.\cite{Ahmady:2016ujw}

\section{CGC DIPOLE MODEL}
In the dipole picture, the scattering amplitude for the diffractive process $\gamma^* p \to V p$ factorizes into an overlap of photon and vector meson light-front wavefunctions and a dipole cross-section \cite{Watt:2007nr}:
\bea
\Im \mbox{m}\, \mathcal{A}_\lambda(s,t;Q^2)  
  = \sum_{h, \bar{h}} \int {\mathrm d}^2 {\mathbf r} \; {\mathrm d} x \; \Psi^{\gamma^*,\lambda}_{h, \bar{h}}(r,x;Q^2)  \Psi^{V,\lambda}_{h, \bar{h}}(r,x)^* e^{-i x \mathrm{r} \cdot \mathbf{\Delta}} \mathcal{N}(x_{\text{m}},\mathrm{r}, \mathbf{\Delta})\; ,
\label{amplitude-VMP} 
\eea
where $t=-\mathbf{\Delta}^2$ is the squared momentum transfer at the proton vertex. $\Psi^{\gamma^*,\lambda}_{h, \bar{h}}(r,x;Q^2)$ and $\Psi^{V,\lambda}_{h, \bar{h}}(r,x)$  are the light-front wavefunctions of photon and vector meson respectively while $\mathcal{N}(x_{\text{m}},\mathrm{r},\mathrm{\Delta})$ is the proton-dipole scattering amplitude. $h$ and $\bar{h}$ are the helicities of the quark and  the antiquark respectively. $r$ is the transverse size of the  color dipole and $x$ is the fraction of light-front momentum of the photon (or vector meson) carried by the quark. Both wavefunctions are labeled by $\lambda=L,T$ which denotes the polarization of the photon or vector meson. The photon light-front wavefunction is also a function  of the photon's virtuality $Q^2$. The dipole-proton  scattering amplitude  is the amplitude for the elastic scattering of the dipole on the proton and it depends on the photon-proton centre-of-mass energy $W$ via the modified Bjorken variable $x_{\mbox{\tiny m}}$ where
\begin{equation}
x_{\text{m}}=x_{\text{Bj}}\left(1+ \frac{M_V^2}{Q^2} \right)~\text{with}~x_{\text{Bj}}=\frac{Q^2}{W^2}
\;.
\label{Bjorken-x}
\end{equation}
The dipole-proton scattering amplitude is a universal object, appearing also in the formula for the fully inclusive DIS process: $\gamma^* p \to X$. In fact, by replacing the vector meson by a virtual photon in Equation \eqref{amplitude-VMP}, we obtain the amplitude for elastic Compton scattering $\gamma^* p \to \gamma^* p$, i.e.
\bea
\left. \Im \mbox{m}\, \mathcal{A}_\lambda(s,t) \right|_{t=0}  
&=&  s \sum_{h, \bar{h}} \int {\mathrm d}^2 {\mathbf r} \; {\mathrm d} x \; |\Psi^{\gamma^*,\lambda}_{h, \bar{h}}(r,x;Q^2)|^2  \hat{\sigma}(x_{\text{m}}, r) \; .
\label{amplitude-compton} 
\eea
$\hat{\sigma}$ in Equation \eqref{amplitude-compton} is the dipole cross-section defined as follows:
\begin{equation}
\hat{\sigma}(x_{\text{m}},r)=\frac{\mathcal{N}(x_{\text{m}},\mathrm{r}, \mathbf{0})}  {s}=\int \mathrm d^2 \mathbf{b}~\mathcal{\tilde{N}}(x_{\text{m}},\mathrm{r}, \mathbf{b}) \;,
\label{dipole-xsec}
\end{equation}
where $\mathcal{\tilde{N}}$ is the Fourier transform of $\mathcal{{N}}$ in the $\mathbf{b}$ (impact parameter) space. Indeed, the elastic amplitude given by Equation \eqref{amplitude-compton} is directly related to the inclusive $\gamma^* p \to X$ total cross-section in DIS via the Optical Theorem: 
\begin{equation}
\sigma_{\lambda}^{\gamma^* p \to X} = \sum_{h, \bar{h},f} \int {\mathrm d}^2 {\mathbf r} \; {\mathrm d} x \; |\Psi^{\gamma^*,\lambda}_{h, \bar{h}}(r,x;Q^2)|^2  \hat{\sigma}(x_{\text{m}}, r)\; ,
\label{gammapxsec}
\end{equation}
where now  \cite{Rezaeian:2013tka}
\begin{equation}
x_{\text{m}}=x_{\text{Bj}}\left(1+ \frac{4m_f^2}{Q^2} \right)~\text{with}~x_{\text{Bj}}=\frac{Q^2}{W^2}
\;.
\label{Bjorken-x-DIS}
\end{equation}

Therefore, one can use the high quality DIS data from HERA to constrain the free parameters of the dipole cross-section section and then use the same dipole cross-section to make predictions for vector meson production.

To lowest order in $\alpha_{\mbox{em}}$, the perturbative photon wavefunctions are given by \cite{Forshaw:2003ki}:
\bea \Psi^{\gamma,L}_{h,\bar{h}}(r,x; Q^2, m_f)  &=& \sqrt{\frac{N_{c}}{4\pi}}\delta_{h,-\bar{h}}e\, e_{f}2 x(1-x) Q \frac{K_{0}(\epsilon r)}{2\pi}\;, 
\label{photonwfL} \\
\Psi^{\gamma,T}_{h,\bar{h}}(r,x; Q^2, m_f) &=& \pm \sqrt{\frac{N_{c}}{2\pi}} e \, e_{f} 
\big[i e^{ \pm i\theta_{r}} (x \delta_{h\pm,\bar{h}\mp} -  (1-x) \delta_{h\mp,\bar{h}\pm}) \partial_{r}   +  m_{f} \delta_{h\pm,\bar{h}\pm} \big]\frac{K_{0}(\epsilon r)}{2\pi}\; , \label{photonwfT}
\eea
where 
$ \epsilon^{2} = x(1-x)Q^{2} + m_{f}^{2} $ and $r e^{i \theta_{r}}$ is the complex notation for the transverse separation between the quark and anti-quark. As is evident from Equation \eqref{photonwfT}, at $Q^2 \to 0$ or $x \to (0,1)$, the photon light-front wavefunctions become sensitive to the non-zero quark mass $m_f$ which prevents the modified Bessel function $K_0(\epsilon r)$ from diverging, i.e. the quark mass acts as an infrared regulator. On the other hand, a non-perturbative model  for the meson light-front wavefunction is used and assumed to be valid for all $r$. 

A simple model for the $b$-integrated dipole-proton amplitude, i.e. the dipole cross-section in Equation \eqref{dipole-xsec} has been proposed in Ref. \cite{Iancu:2003ge}. This is known as the CGC dipole model and is given by
\be 
\hat{\sigma}(x_{\text{m}},r) = \sigma_0 \, {{ \mathcal N } (x_{\text{m}}, r Q_s, 0 ) }\; ,
\ee
with \bea
{ \mathcal N } (x_{\text{m}},r Q_s, 0) &=& { \mathcal N}_0 \left ( { r Q_s \over 2 }\right)^ {2 \left [ \gamma_s + { {\mathrm ln}  (2 / r Q_s) \over  \kappa \, \lambda \, {\mathrm ln} (1/x_{\text{m}}) }\right]}  ~~~~~~~{\rm for } ~~~~~ ~~~~ r Q_s \leq 2 \nonumber \\
&=& { 1- \exp[-{\mathcal A} \,{\mathrm ln}^2 ( {\mathcal B} \, r Q_s)]}  ~~~~~~~~~{\rm for } ~~~~~ ~~~~ r Q_s > 2 \; ,
\eea
where the saturation scale $Q_s = (x_0/x_{\text{m}})^{\lambda / 2}$ GeV. The coefficients ${\mathcal A}$  and ${\mathcal B}$ are determined  from the condition that the ${\mathcal N}(r Q_s, x)$ and its derivative with respect to $r Q_s$ are  continuous at $r Q_s=2$.  This leads to
\be  {\mathcal A} =  - { ({ \mathcal N}_0 \gamma_s)^2  \over (1 - { \mathcal N}_0)^2 \, {\mathrm ln}[1 - { \mathcal N}_0] }\,, ~~~~~~
{\mathcal B} = {1 \over 2} (1 - { \mathcal N}_0)^{-{(1 - { \mathcal N}_0) \over { \mathcal N}_0 \gamma_s }}\,.
\ee
The free parameters of the CGC dipole model are $\sigma_0, \lambda, x_0$ and $\gamma_s$ which are fixed by a fit to the structure function $F_2$ data.  In 2015, the H1 and ZEUS collaborations have released highly precise combined data sets \cite{Abramowicz:2015mha} for the reduced cross-section 
\begin{equation}
\sigma_r(Q^2,x,y)=F_2(Q^2,x)-\frac{y^2}{1+(1-y)^2} F_L(Q^2,x) \; ,
\label{sigmar}
\end{equation}
where $y=Q^2/\hat{s}x$ and $\sqrt{\hat{s}}$ is the centre of mass energy of the $e p$ system for $4$ different bins : $\sqrt{\hat{s}}=225$ GeV ($78$ data points),  $\sqrt{\hat{s}}=251$ GeV ($118$ data points) and $\sqrt{\hat{s}}=300$ GeV ($71$ data points), $\sqrt{\hat{s}}=318$ GeV ($245$ data points).  Our fitted values for the CGC dipole model parameters  together with the resulting  $\chi^2$ per degrees of freedom ($\chi^2/\mbox{d.o.f}$) values are shown in Table \ref{tab:F2fit}. The first two rows indicate that the fit is not very sensitive to the variation in the strange quark mass. Comparing the second and third rows, we can see that the data prefer the lower $u$ and $d$ quark masses and that increasing them give quite different fit parameters especially for $x_0$.

%%%%%%%%%%%%%%%%%%%%%%%%%%%%%%%%%
\begin{table}[h]
	\caption{Parameters of the CGC dipole model extracted from our fits to inclusive DIS data (with $x_{\text{Bj}} \le 0.01$ and $Q^2\in [0.045, 45]\,\text{GeV}^2$) using $3$ different sets of quark masses.}
	\label{tab:F2fit}
	\tabcolsep7pt
	\begin{tabular}{lcccccc}
		\hline
		 \tch{1}{c}{b}{$[m_{u,d},m_s]$/GeV}  & \tch{1}{c}{b}{$\gamma_s$} & \tch{1}{c}{b}{$\sigma_0$/mb}  & \tch{1}{c}{b}{$x_0$}  & \tch{1}{c}{b}{$\lambda$} & \tch{1}{c}{b}{$\chi^2/\mbox{d.o.f}$}  \\
		\hline
			$ [0.046,0.357]$ & $0.741$  &  $26.3$ & $1.81 \times 10^{-5}$ & $0.219$ & 535/520=1.03\\
			$ [0.046,0.14]$ & $0.722$  &  $24.9$ & $1.80 \times 10^{-5}$ & $0.222$ & 529/520=1.02\\
			$ [0.14,0.14]$ & $0.724$  &  $29.9$ & $6.33 \times 10^{-6}$ & $0.206$ & 554/520=1.07\\
		\hline
	\end{tabular}
\end{table}

% Head 2
\section{HOLOGRAPHIC MESON LFWF}
\label{Holographic wfn}
The vector meson light-front wavefunctions appearing in Equation \eqref{amplitude-VMP} cannot be computed in perturbation theory. Explicitly, the vector meson light-front wavefunctions can be written as \cite{Forshaw:2012im}
\be
\Psi^{V,L}_{h,\bar{h}}(r,x) =  \frac{1}{2} \delta_{h,-\bar{h}}  \bigg[ 1 + 
{ m_{f}^{2} -  \nabla_r^{2}  \over x(1-x)M^2_{V} } \bigg] \Psi_L(r,x) 
\label{mesonL}
\ee
and
\be \Psi^{V, T}_{h,\bar{h}}(r,x) = \pm \bigg[  i e^{\pm i\theta_{r}}  ( x \delta_{h\pm,\bar{h}\mp} - (1-x)  \delta_{h\mp,\bar{h}\pm})  \partial_{r}+ m_{f}\delta_{h\pm,\bar{h}\pm} \bigg] {\Psi_T(r,x) \over 2 x (1-x)}\,. 
\label{mesonT}
\ee

Various ansatz for the non-perturbative meson wavefunction have been proposed in the literature, but in recent years, new insights about hadronic light-front wavefunctions based on the anti-de Sitter/Conformal Field Theory (AdS/CFT) correspondence have been proposed by Brodsky and de T\'eramond. \cite{Brodsky:2014yha}. In this framework the vector meson wavefunctions $\Psi_\lambda(r,x),\; \lambda=T,\; L$ are given as
\be  \Psi_{\lambda} (x,\zeta) = \mathcal{N}_{\lambda} \sqrt{x (1-x)}  \exp{ \left[ -{ \kappa^2 \zeta^2  \over 2} \right] }
\exp{ \left[ -{m_f^2 \over 2 \kappa^2 x(1-x) } \right]} \; ,
\label{hwf}
\ee
where we have introduced a polarization-dependent normalization constant ${\mathcal N}_{\lambda}$. We fix this normalization constant by requiring that
\be
\sum_{h,\bar{h}} \int {\mathrm d}^2 {\mathbf{r}} \, {\mathrm d} x |
\Psi^{V, \lambda} _{h, {\bar h}}(x, r)|^{2} = 1 \; ,
%\label{normalisation}
\ee
where $\Psi^{V,\lambda}_{h, \bar {h}}(x,r)$ are given by Equations \eqref{mesonL} and \eqref{mesonT}. 

\section{COMPARISON WITH DATA}
Having specified the dipole cross-section and the holographic meson wavefunction, we can now compute cross-section for diffractive $\phi$ production.  We shall show predictions using three sets of the CGC dipole parameters as given in Table \ref{tab:F2fit}. We shall refer to these three sets of predictions as Fit A (first row), Fit B (second row) and Fit C (third row) respectively. Recall that all our predictions will be generated using the same holographic wavefunction given by Equation \eqref{hwf} and they differ only by the choice of quark masses and the corresponding fitted parameters of the CGC dipole model as given in Table \ref{tab:F2fit}. 
 
\begin{figure}
	\centerline {%
		\includegraphics
		[width =.55\textwidth ]{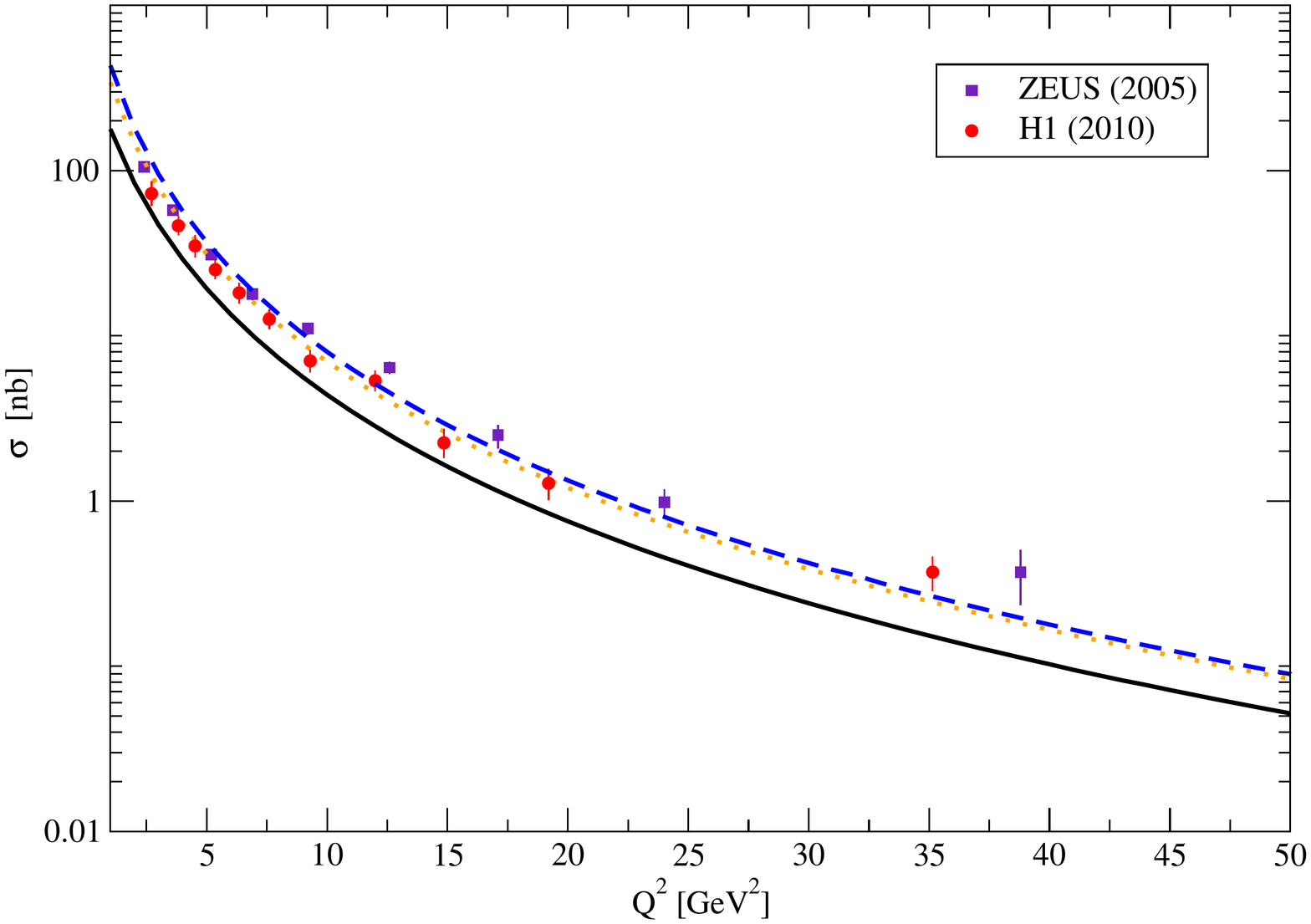}\hskip -0.75cm\includegraphics[width=.55\textwidth]{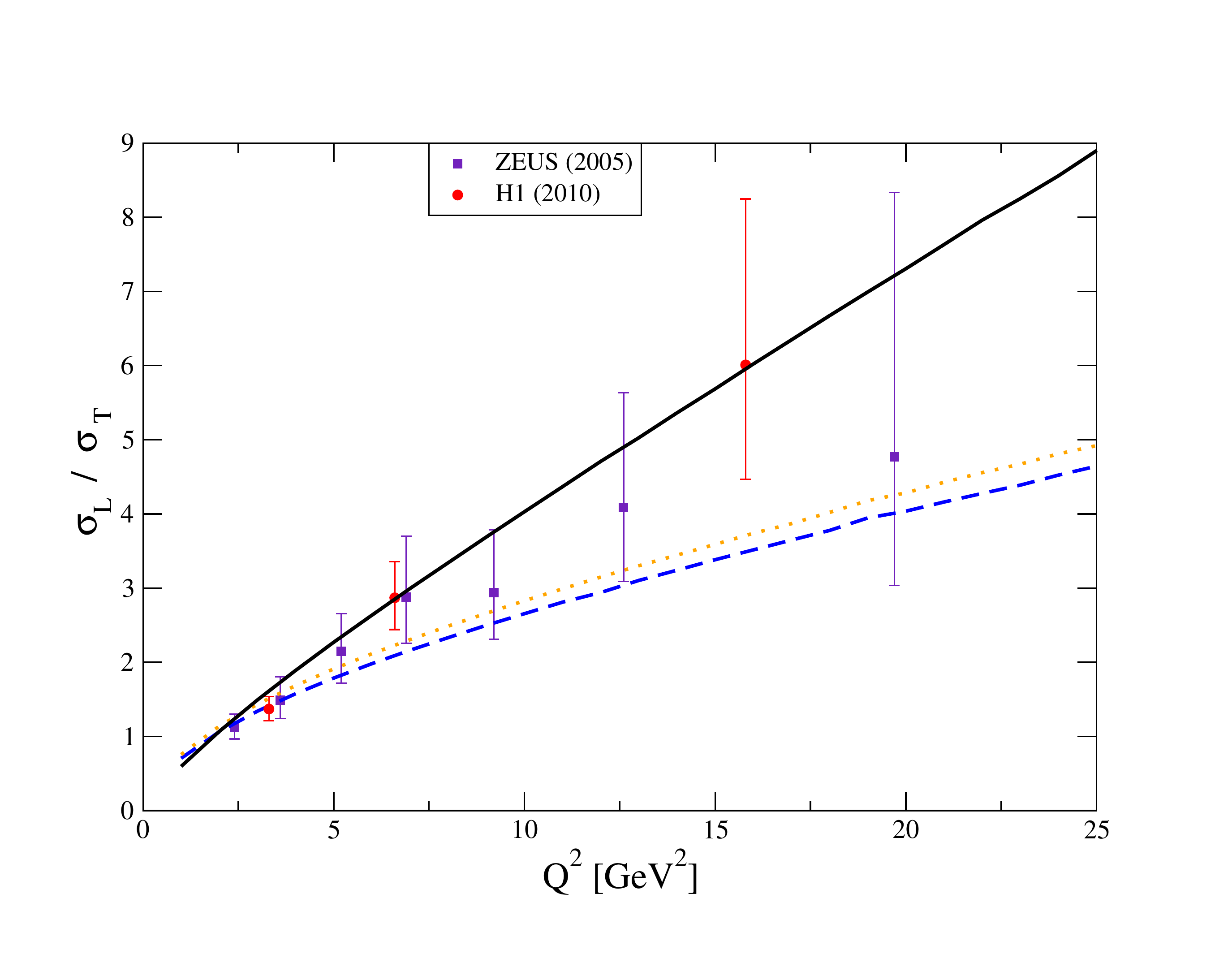}}
	\caption{Predictions for the $\phi$ production total cross section (left) and longitudinal to transverse cross-section ratio (right) at $W=90$ GeV as a function of $Q^2$  compared to HERA data \cite{Chekanov:2005cqa,Aaron:2009xp}. Black solid curve: A. Orange dotted curve: B. Blue dashed curve: C.}
	\label{phi-q2}
	\end{figure}

For $\phi$ production, our predictions for the $Q^2$ dependence of the total cross-section at fixed $W$ are shown in Figure~\ref{phi-q2}. Here, it is clear that the Fit A predictions (solid black curves) are not successful. The data prefer slightly the Fit B (orange dotted curves) over the Fit C predictions (blue dashed curves) although the lack of data in the low $Q^2$ region prevents us from making a definite statement. At high $Q^2$, our predictions tend to undershoot the (ZEUS) data as expected. Our predictions for the longitudinal to transverse cross-sections ratio for $\phi$ production are shown in Figure~\ref{phi-q2}. We can see that the ratio data tend to favour the Fit A  prediction (solid black curve) although they are not precise enough to discard the other two predictions.

%\section{CONCLUSION}
%The holographic light-front meson wavefunction is successful in describing diffractive $\phi$ production with a suitable choice of light quark masses.
% Acknowledgement
\section{ACKNOWLEDGMENTS}
This research is supported by a Team Discovery Grant from the Natural Sciences and Engineering Research Council (NSERC) of Canada.

% References

%\nocite{*}
\bibliographystyle{aipnum-cp}%
\bibliography{ahmady}%

\end{document}